\documentclass{vgtc}                          % final (conference style)
\ifpdf%                                % if we use pdflatex
  \pdfoutput=1\relax                   % create PDFs from pdfLaTeX
  \usepackage{graphicx}                % allow us to embed graphics files
  \DeclareGraphicsExtensions{.pdf,.png,.jpg,.jpeg} % for pdflatex we expect .pdf, .png, or .jpg files
\else%                                 % else we use pure latex
  \usepackage{graphicx}                % allow us to embed graphics files
  \DeclareGraphicsExtensions{.eps}     % for pure latex we expect eps files
\fi%

\graphicspath{{figures/}{pictures/}{images/}{./}} % where to search for the images

\usepackage{microtype}                 % use micro-typography (slightly more compact, better to read)
\PassOptionsToPackage{warn}{textcomp}  % to address font issues with \textrightarrow
\usepackage{textcomp}                  % use better special symbols
\usepackage{mathptmx}                  % use matching math font
\usepackage{times}                     % we use Times as the main font
\usepackage{cite}                      % needed to automatically sort the references
\usepackage{tabu}                      % only used for the table example
\usepackage{booktabs}   
\usepackage{subfig}
\usepackage{verbatim}
\usepackage[dvipsnames]{xcolor}
\usepackage{soul}
\usepackage{hyperref}
\onlineid{0}

\vgtccategory{Research}

\vgtcinsertpkg

\title{TempoCave: Visualizing Dynamic Connectome Datasets \\to Support Cognitive Behavioral Therapy}

\author{Ran Xu$^1$, Manu Mathew Thomas$^1$, Alex Leow$^2$, Olusola A. Ajilore$^2$, Angus G. Forbes$^1$ \\ \\
\parbox{2.4in}{\scriptsize \centering $^1$Department of Computational Media \\ University of California, Santa Cruz \\ \{rxu3,mthomas6,angus\}@ucsc.edu}
\parbox{2.4in}{\scriptsize \centering $^2$Department of Psychiatry\\ University of Illinois at Chicago \\ \{weihliao,oajilore\}@uic.edu}}

\teaser{
\centering
  \includegraphics[width=6.25in]{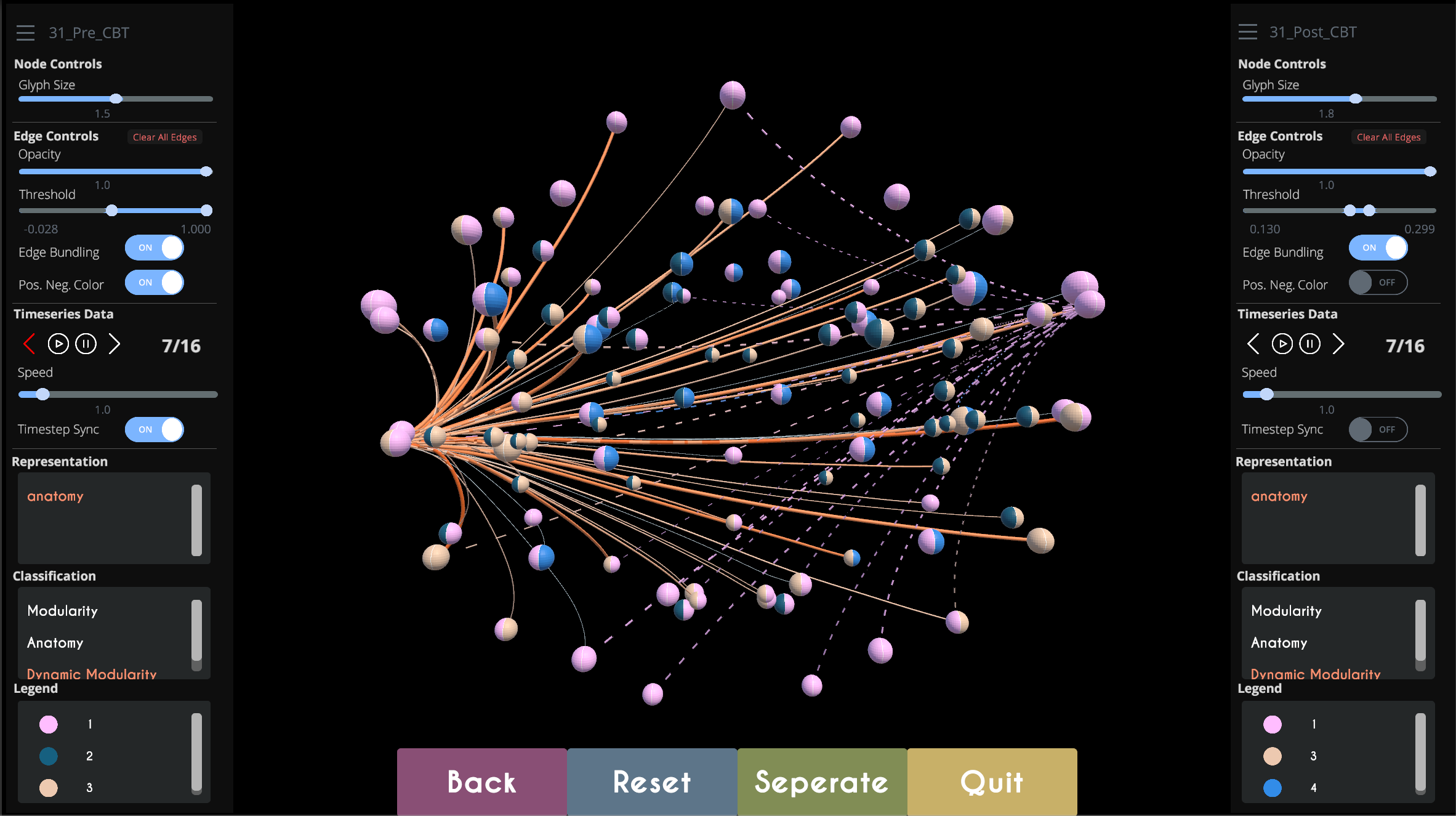}
  \caption{
  A screenshot of the \textit{TempoCave} application. Here, a user compares frames from pre- and post-treatment dynamic connectomes for an individual patient with major depression disorder. Using the option panels on either side of the application, a user can choose different visual encodings to accentuate features useful for understanding the activity of particular brain regions. The left and right coloring of the nodes indicates the modular affinity of a brain region for a patient's pre- and post-treatment connectome, respectively. Likewise, the styling of the connections indicates either the connectome they belong to or the strength of the correlation. A user can synchronize playback of the frames of the connectomes or compare selected frames on demand to gain insight into their dynamics, and during an analysis session a user can interactively toggle on or off brain regions of interest, or switch to alternative representations of the connectome defined  using layouts based on dimensionality reduction techniques. 
  }
  \label{fig:teaser-image}
}

\abstract{
We introduce \textit{TempoCave}, a novel visualization application for analyzing dynamic brain networks, or connectomes. \textit{TempoCave} provides a range of functionality to explore metrics related to the activity patterns and modular affiliations of different regions in the brain. These patterns are calculated by processing raw data retrieved functional magnetic resonance imaging (fMRI) scans, which creates a network of weighted edges between each brain region, where the weight indicates how likely these regions are to activate synchronously. In particular, we support the analysis needs of clinical psychologists, who examine these modular affiliations and weighted edges and their temporal dynamics, utilizing them to understand relationships between neurological disorders and brain activity, which could have significant impact on the way in which patients are diagnosed and treated. We summarize the core functionality of \textit{TempoCave}, which supports a range of comparative tasks, and runs both in a desktop mode and in an immersive mode. Furthermore, we present a real world use case that analyzes pre- and post-treatment connectome datasets from 27 subjects in a clinical study investigating the use of cognitive behavior therapy to treat major depression disorder, indicating that \textit{TempoCave} can provide new insight into the dynamic behavior of the human brain. } % end of abstract
\begin{document}
\firstsection{Introduction}

\maketitle %AGF - this needs to be AFTER the \firstsection command
\label{sec:intro-tasks}

\begin{figure*}[h!]
\centering
\includegraphics[width=7in]{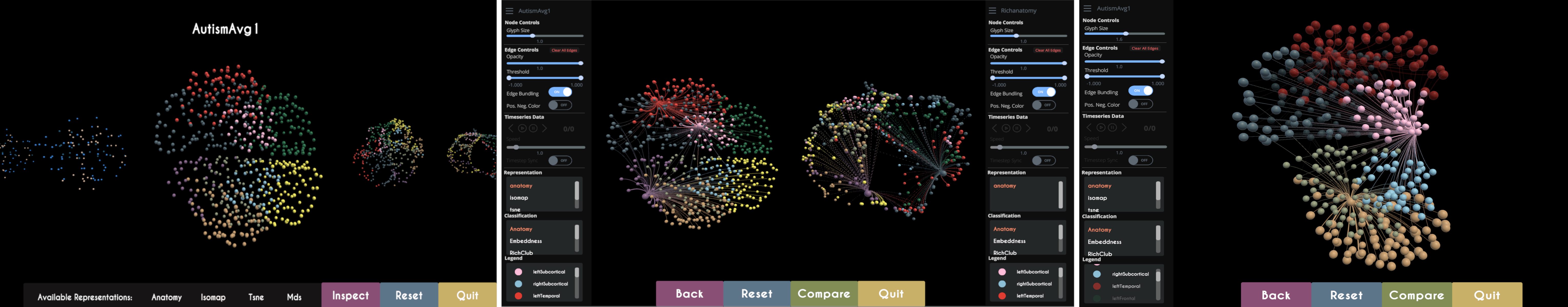}
\caption{Different views in the \textit{TempoCave} interface. Upon starting the application, a thumbnail of all connectomes in the data folder appear in a carousel view (left). Users can add connectomes to the inspection view (middle), which provides a menu panel for each of the selected connectomes and enables users to interactively investigate various connectome features. Users can zoom into a single connectome (right) to more clearly examine the modular affiliation of specific brain regions and to investigate the strength of connections between these regions on demand. In the inspection view, as well as the comparison view shown in Fig.~\ref{fig:teaser-image}, a user can also interactively rotate the connectome, toggle edge bundling on or off, change the transparency of edges, increase the size of the nodes, change the visual encodings for nodes and edges, and show or hide selected brain regions or communities. If the connectome has associated alternative layouts, then the user can switch between these views as well, and if the connectome is dynamic, then user can play through the data, or jump to a specified time frame.}
\label{fig:application}
\end{figure*}

The human brain is comprised of trillions of structured neurophysiological connections. Clinical neuroscientists create \textit{connectome} representations of this vastly complex biological network via functional magnetic resonance imagine (fMRI) scans~\cite{logothetis2001neurophysiological}, producing a map of the connectivity between cortical and subcortical brain regions. Understanding the structure and functionality of the connectome is a fundamental goal of cognitive neuroscience and neuropsychology, and provides insight into the behavior of various brain conditions~\cite{sporns2005human}. Even when displaying the (relatively) low-resolution network representations generated by non-invasive neuroimaging techniques (where a node represents a brain region ``voxel'' containing millions of neurons), the visual complexity can be overwhelming. Nonetheless, this network of nodes enables graph-theoretic methods that extract useful connectome metrics, and the edges linking the nodes provide finer details about the strength of the connectivity between regions. This complexity is exacerbated when analyzing dynamic connectomes~\cite{cooper2018time,preti2017dynamic}, in which a user investigates temporal patterns across the sequence of frames generated during a dynamic fMRI scanning session. Moreover, comparing two (or more) connectomes is necessary when evaluating pre- vs. post-treatment changes~\cite{carhart2017psilocybin}, negative vs. positive connections~\cite{zhanimportance}, or healthy vs. unhealthy connectomes~\cite{ramakrishna2018functional}, which introduces additional visualization challenges.

In this paper, we introduce \textit{TempoCave}, a visualization application that facilitates the exploration and analysis of time-series connectome datasets. We worked closely with our neuroscience collaborators to identify aspects of their analysis workflow not yet supported by existing tools, and to determine which tasks are most relevant for making sense of dynamic connectomes in clinical contexts:
\vspace{2mm}

\setlength{\parindent}{.5em}
\textbf{T1}: Measuring the \textit{\underline{dwelling time}} of brain regions, defined as the length of time a region spends in the dominant community.

\textbf{T2}: Measuring the \textit{\underline{flexibility}} or particular brain regions, defined as the number of times a region changes modular affiliation. 

\textbf{T3}: Analyzing the \textit{\underline{connectivity}} between brain regions that is used to define modular affiliations, including edges that are defined by either a negative or positive correlation between nodes. 

\textbf{T4}: Understanding the overall \textit{\underline{dynamics}}, or ``stickiness'' of the connectome, in terms of how dwelling, flexibility, and connectivity metrics change over time.

\textbf{T5}: Enabling the \textit{\underline{comparison}} of multiple dynamic connectomes, for example, comparing a connectome to a group average, or comparing an individual patient's connectome pre- vs. post-treatment.

\vspace{2mm}

\setlength{\parindent}{1em}

\noindent These tasks have broad relevance to psychiatry, as a wide range of neurological disorders are linked to the disruption of normal brain connectivity. An important overarching goal of clinical neuroscience is to find relationships between brain activity and neurological disorders, which can then be leveraged to make diagnoses, to guide treatment plans, and to better understand the human brain. We developed \textit{TempoCave} iteratively, adding relevant visualization features as we worked successively with individual patients' connectomes and group average connectomes for autism spectrum disorder, anxiety disorder, and major depressive disorder. In this paper, we focus on connectome datasets gathered during a clinical study on \textit{rumination}, a mental disorder characterized by repetitively and passively focusing on symptoms of distress and its causes. The consequences of prolonged rumination include anxiety and depression~\cite{berman2010depression,kucyi2015dynamic,wang2015individual}. 

Rumination-focused cognitive behavior therapy, or R-CBT, assists individuals in realizing that their rumination about negative experience can be unhelpful, and coaches them on how to shift to a more helpful style of thinking. For example, patients undergoing R-CBT are asked to remember previous positive mental states, such as a time they were completely absorbed in an activity--- the opposite of ruminating~\cite{koster2011understanding,watkins2007rumination}. R-CBT appears to be effective at supporting emotion regulation in patients suffering from depression. A preliminary study (described in more detail in Sec.~\ref{s:use-case}) finds that patients who continued treatment remained in remission after 8 weeks, whereas patients who did not continue treatment had a higher likelihood of relapse. \textit{TempoCave} facilitates insight into how R-CBT (and other clinical interventions) alters brain behavior. In order to conduct these analyses, clinical neuroscientists make detailed measurements of the brain activity of patients and compare the patients' connectomes before treatment and after treatment, and also compare them against baseline healthy connectomes. Visualizing the dynamics of connections within and between brain regions, measured by dwelling time and flexibility, helps clinical psychiatrists examine the ``stickiness'' of these connections, where comparatively high measurements of stickiness can imply an unhealthy connectome. 

To our best knowledge, \textit{TempoCave} is the first application to support comparison tasks for multiple time-series connectome networks in order to better understand rumination. Our contributions include: (a) a delineation of analysis tasks relevant to reasoning about dynamic connectome datasets; (b) the introduction of a new visualization tool to support connectome-based comparison tasks for both static and dynamic networks; (c) techniques for observing and analyzing community affiliation of nodes in a dynamic network; (d) an analysis pipeline that supports easy data loading of multiple connectomes, including those in alternative topological spaces~\cite{allen2015intrinsic} or generated using different modularity identification algorithms~\cite{gadelkarim2014investigating,zhanimportance}; (e) a real-world use case illustrating how \textit{TempoCave} is used in a clinical setting to elucidate new insight into neurological aspects of rumination. Fig.~\ref{fig:teaser-image} presents an overview of \textit{TempoCave} comparing two dynamic connectomes.

\begin{figure*}[t!]
\includegraphics[width=7in]{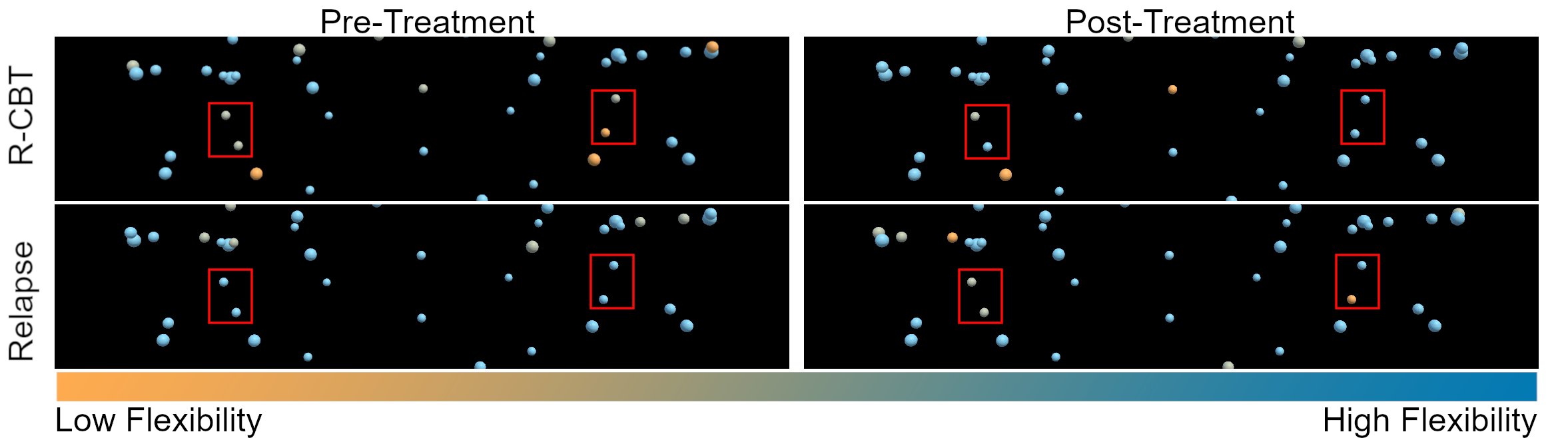}
\caption{This figure shows a summary view of the \textit{flexibility} of brain regions associated with rumination (\textbf{T2}). The top images show the connectome of a MDD patient who received R-CBT treatment, indicating higher flexibility (blue nodes) post-treatment than pre-treatment (where the orange nodes indicate lower flexibility). The bottom images show a relapsed patient who did not receive R-CBT, where there is substantially decreased flexibility in brain regions associated with rumination. The red rectangle highlights the two relevant brain regions: supramarginal gyrus (top) and angular gyrus (bottom).}
\label{Fig:use-case}
\end{figure*}

\section{Background \& Related Work}
\textit{TempoCave} is inspired by previous approaches for visualizing networks~\cite{beck2017taxonomyCGF}, comparing graphs~\cite{gleicher2011visual}, and exploring connectome datasets~\cite{margulies2013visualizing}. Keiriz et al.~\cite{keiriz2018neurocave} survey the landscape of connectome visualization, focusing mainly on static datasets, and characterizing them in terms of their main visualization modality (emphasizing volume, surface, or graph representations), as well as identifying tools that include support for virtual reality (VR) environments. \textit{TempoCave} presents connectomes as 3D networks, similar to approaches presented in \textit{Connectome Visualization Utility}~\cite{laplante2014connectome}, \textit{BrainNet Viewer}~\cite{xia2013brainnet}, \textit{Connectome Viewer Toolkit}~\cite{gerhard2011connectome}, and the \textit{AlloBrain} project~\cite{thompson2009allobrain}. However, \textit{TempoCave} specifically focuses on visualizing nodes and edges to make it easier to reason about metrics associated with modular affiliation and connectivity.

\textit{TempoCave} provides features for dynamic data analysis, presenting a synchronized playback mode that highlights differences between two connectomes at different points in time. A number of 2D visualizations have been used to explore dynamic connectome data~\cite{hutchison2013resting, allen2012tracking, ma2015visualizing, de2017temporaltracks}, and Beck et al.~\cite{beck2017taxonomyCGF} summarize approaches to visualize 2D dynamic networks. Other tools instead utilize a 3D layout for investigating dynamic connectome data~\cite{he2011econnectome,liao2014dynamicbc,Ma2015_JIST_SwordPlots}. For example, Xing et al.'s \textit{Thought Chart}~\cite{xing2016thought} presents distinct 3D trajectories for different task conditions and provides a comparative analysis that generates a summary view of how much one dynamic connectome differs in comparison to others. Similar to our implementation, Arsiwalla et al. introduce \textit{BrainX$^3$}\cite{arsiwalla2015network}, an interactive and immersive 3D visualization of dynamic connectomes, but which does not support comparison tasks, a main feature of \textit{TempoCave}. 

Alper et al.~\cite{alper2013weighted} investigate visual encodings of edge weights within an adjacency matrix to support comparisons of brain connectivity patterns. \textit{TempoCave} further emphasizes comparisons of modularity affiliation metrics that summarize the edge weights in the network. Other recent brain visualization tools focus on neurobiological tasks, such as Ganglberger et al.'s~\cite{ganglberger2019braintrawler} \textit{BrainTrawler}, which provides tools to conduct integrated analyses of genomic data and mesoscale neuroscience datasets at the level of individual neurons. \textit{TempoCave} aims to support tasks \textbf{T1}-\textbf{T5} relevant to diagnosing and treating patients with neurological disorders, extending previous work that focused on static connectomes~\cite{keiriz2018neurocave}, which also presents a 3D view and enables a user to explore data immersively~\cite{conte2015braintrinsic,forbes2014stereoica}.

\section{The \textit{TempoCave} Application}
\label{sec:Application}

\textit{TempoCave} is an interactive tool that enables clinical psychiatrists to load, visualize, and analyze dynamic connectomes, solving several technical challenges associated with the analysis tasks described in Sec.~\ref{sec:intro-tasks}. \textit{TempoCave} supports the investigation of multiple connectomes at once, superimposing connectomes to make comparisons, and providing details about modular affiliation and edge dynamics that are needed in order to understand some dynamic datasets.

\label{ss:data-acquisition}
\begin{comment}
\begin{figure*}
\subfloat[Selection View]{\includegraphics[width = 1in,height = 0.6in]{Images/FirstScene.PNG}}
\subfloat[Inspection View]{\includegraphics[width = 1in, height = 0.6in]{Images/SecondScene.PNG}}
\subfloat[Pos. vs Neg. Connect]{\includegraphics[width = 1in, height = 0.6in]{Images/21_post_PosNeg.PNG}}
\caption{}
\end{figure*}
\end{comment}

Our clinical neuroscience collaborators capture connectome datasets using high-resolution functional magnetic resonance imaging (fMRI) scans. The fMRI data is pre-processed to extract node and edge information, such as connection strength, and to perform various dimensional reduction steps. Dynamic datasets are obtained using scans taken at regular intervals over a short period of time. In the clinical study presented in Sec.~\ref{s:use-case}, 200 frames are captured over the course of a $\sim$6 minute scanning session, and then further processed using the PACE algorithm~\cite{zhanimportance} to determine modular affiliations.

\begin{figure*}[t!]
\centering
\includegraphics[width=7in]{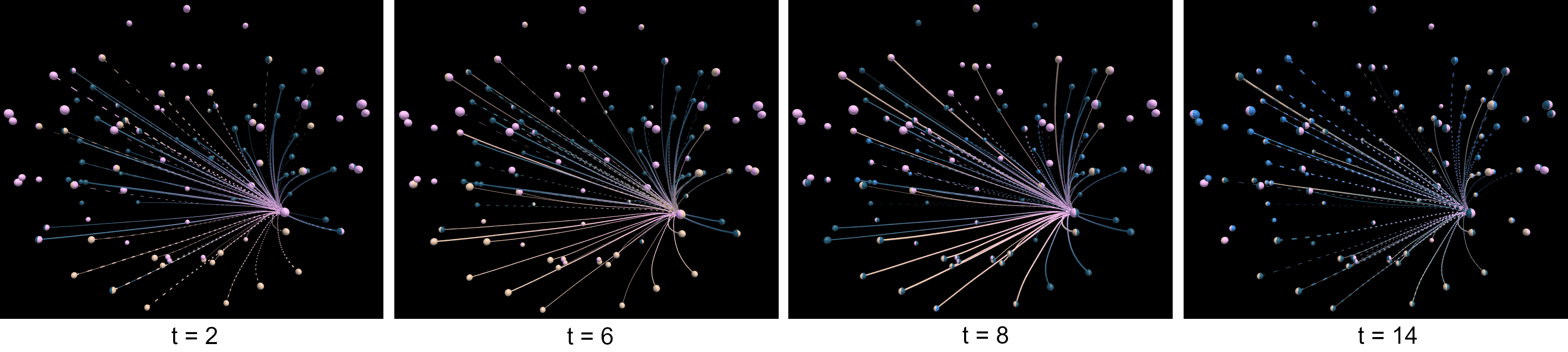}
\caption{This figure presents the dynamic connectome of a patient with MDD who received R-CBT. Here, the overlay comparison reveals differences in modular affiliation and in connectivity patterns pre- vs. post-treatment across multiple time steps. }
\label{fig:Time-series}
\end{figure*}

%\subsection{User Interface Design}
\textit{TempoCave} consists of a selection interface (Fig.~\ref{fig:application} left), and an interactive inspection interface (Fig.~\ref{fig:application} middle and right). The selection interface displays show an overview ``carousel'' of the available connectomes, along with a list of all the available layouts (generated through different dimension reduction algorithms, such as Isomap or t-SNE). The user can select two (in desktop mode) or more (in VR mode) connectomes for analysis and comparison. The inspection interface shows the selected connectomes with each connectome having their own settings panel, which provides options: to change the representation of the connectomes, to update the classification of different brain regions, to filter edges based on their connection strength, and to toggle edge bundling to mitigate visual clutter. 

%\subsection{Overview Visualization}
Summarizing the modular affiliation or ``stickiness'' of the nodes in dynamic connectomes is an important aspect of analyzing a patient's connectome, as is investigating the amount of time that brain regions are associated with particular communities (dwelling time) and the frequency that brain regions change affiliation (flexibility). Connectomes demonstrating patterns of more rapid modular change are believed to indicate healthier connectomes. \textit{TempoCave} automatically processes the summary statistics for each dynamic connectome upon being loaded into the application. For each node in the connectome, changes in the affiliation across the time steps are used to determine the flexibility metric. Dwelling time is calculated by identifying the maximum time a node is associated with a specific module. \textit{TempoCave} uses color-coding to represent both dwelling time and flexibility, supporting \textbf{T1} and \textbf{T2}. Fig.~\ref{Fig:use-case} shows how \textit{TempoCave} is used to analyze flexibility in a rumination study.   

%\subsection{Dynamic Visualization}
To analyze changes in dwelling time, flexibility, and edge connectivity over time, \textit{TempoCave} provides controls to scrub through the time steps. Users can play and pause, move forward or backward one or more time steps, adjust the playback speed, and, if two connectomes are available for inspection, there is an option to synchronize the playback settings. At each time step nodes are colored based on their modular affiliation. The edges change their width based on the strength of the connectivity between two nodes, and are colored with a gradient representing the regions they are associated with. The edges can further be classified into positive connections if the regions are correlated or negative connections if they are uncorrelated. \textit{TempoCave} supports an optional color-coding to show the negative and positive edges for each time step, as shown in Fig.~\ref{fig:teaser-image}. The dynamic visualization features support tasks \textbf{T3} and \textbf{T4}. Fig.~\ref{fig:Time-series} shows an example of how the edge connectivity and modularity change over time in an analysis session using \textit{TempoCave}.

\textit{TempoCave} presents an overlay comparison mode supporting \textbf{T5}, where two connectomes are juxtaposed to form an integrated layout. Each node in this superimposed view is split into two halves, corresponding to left and right connectomes. As the user moves through the different frames, each half of the node's colors change separately based on the modular affiliation of the associated connectome. The edges can be activated by clicking the corrresponding half of the node. To distinguish the connectivity of two connectomes, we use a solid line for left connectome and a dashed line for the right connectome. A comparison of a pre- vs post-treatment connectome is shown in Fig.~\ref{fig:teaser-image}.

We developed \textit{TempoCave} using the Unity Engine, which renders 3D data at real-time frame rates and provides out-of-the-box solutions for immersive applications. We evaluated a range of visual encodings and interaction modalities to determine useful representations and interactions that supported the analysis of dynamic connectomes. For example, we initially included animation as a primary encoding for dynamic network features~\cite{romat2018animated}. While our collaborators found the animation engaging, ultimately it was distracting and introduced visual fatigue when they needed to scrub through many time steps. Instead, we color-coded edges to represent both modular affiliation and edge weights, giving the users the option to choose which encoding was most useful for a particular analysis session. We also experimented with a wide range of shapes to represent brain regions, hoping to more clearly distinguish nodes from different connectomes in the comparison view. However, our users found that it was easier to interpret a node when rendered as a multi-colored sphere. Contrary to expectations, we discovered that our users preferred to have fewer available encodings overall, but more options to control which data elements these encodings represent.

\section{Use Case: Investigating R-CBT Treatment}
\label{s:use-case}

As an initial validation of \textit{TempoCave}, we explored a dataset from an ongoing clinical study that measures the effectiveness of rumination cognitive behavioral therapy (R-CBT) for adolescent patients with at least one previous episode of major depression disorder (MDD)~\cite{burkhouse2017neural}. In this study, 27 patients were recruited for comparing 8 weeks of R-CBT data against a control group of 15 healthy participants. The patients were in remission at the start of the study, and the main goal of the study was to measure the effectiveness of R-CBT at preventing a relapse of depression. Of the 27 patients with MDD, 14 were administered a treatment of R-CBT, and none had a relapse during that time. Of the 13 who were not administered R-CBT, 4 patients had an episode of MDD, providing initial evidence of the efficacy of R-CBT.

Using \textit{TempoCave}, clinical psychiatrists observed the modularity dynamics of individuals' connectomes. Participants from a healthy control group were found to have a higher overall flexibility than patients with MDD, and that on average there is no significant difference between the pre- and post-treatment connectomes of MDD patients who received R-CBT. However, as depicted in Fig.~\ref{Fig:use-case}, the post-treatment connectome of MDD patients who did \textit{not} receive R-CBT (and relapsed) shows much less flexibility than was observed in their pre-treatment connectome. Fig.~\ref{Fig:use-case} highlights the supramarginal gyrus and angular gyrus, two regions that are indicated in depression disorder.  Fig.~\ref{fig:Time-series} shows an overlay comparison view of a patient's pre- and post-treatment connectomes who received R-CBT, where the angular gyrus is selected. Looking at the right half of the selected node (from the post-treatment connectome), clinical psychiatrists found that the angular gyrus changed modularity three times (from t=2 to t=14), but that in left half of the selected node (from the pre-treatment connectome) the modularity of angular gyrus changed only once (from t=8 to t=14). This visual analysis corroborates the hypothesis that R-CBT mitigates rumination, as indicated by the temporal dynamics of dwelling time and flexibility metrics.

\section{Conclusion \& Future Work}

Our initial use case already indicates that \textit{TempoCave} helps clinical neuroscientists form new hypotheses about dynamic connectome datasets, and in particular that the comparison mode is useful for providing insight into patterns that emerge when investigating a patient's response to treatment. Future work will explore a wider range of use cases in various clinical contexts. Additional definitions of modularity could generate network metrics that may be useful for understanding brain dynamics. For instance, recent work by Kim and Lee~\cite{kim2019relational} introduce an ``inconsistency'' metric which can be used as an alternative definition of node centrality. Furthermore, while \textit{TempoCave} provides textual labelling of nodes and edges, our collaborators indicated the need for including additional annotation options, which would make it easier to share results or hypotheses with other clinicians and to include as figures in presentations and articles. We also plan to extend our approach to other (non-connectome) dynamic datasets, and to conduct user studies to determine the effectiveness of our visual encodings as a more general approach for highlighting dynamic network features and comparing networks. \textit{TempoCave} is available via our open source GitHub code repository at \url{https://github.com/CreativeCodingLab/TempoCave}, along with source code, detailed instructions on how to load in custom datasets, and additional documentation.

\bibliographystyle{abbrv-doi} 

\bibliography{TempoCave.bib}
\end{document}